\documentclass[aps,pra,twocolumn,superscriptaddress,eps2pdf,notitlepage,10pt]{revtex4-2}

\usepackage[utf8]{inputenc}

\usepackage{hyperref}
\usepackage{graphicx}
\usepackage{amsmath}
\usepackage{amssymb, dsfont}
\usepackage{cleveref}
\usepackage{braket}
\usepackage{enumerate}
\makeatletter
\let\newfloat\newfloat@ltx
\makeatother

\crefname{table}{table}{tables}
\Crefname{table}{Table}{Tables}
\crefname{figure}{Fig.}{Figs.}
\Crefname{figure}{Fig.}{Figs.}
\crefname{equation}{Eq.}{Eqs.}
\Crefname{Equation}{Equation}{Equation}

\newcommand\ns{\mathrm{ns}}

\usepackage{booktabs} 

\usepackage{chngcntr}
\newcounter{lastequationbeforesection}

\begin{document}

\title{Reducing the error rate of a superconducting logical qubit using analog readout information}

\author{Hany Ali}
\affiliation{QuTech and Kavli Institute of Nanoscience, Delft University of Technology, P.O. Box 5046, 2600 GA Delft, The Netherlands}
\affiliation{Present address: Quantware B.V., Elektronicaweg 10, 2628 XG Delft, The Netherlands}

\author{Jorge Marques}
\affiliation{QuTech and Kavli Institute of Nanoscience, Delft University of Technology, P.O. Box 5046, 2600 GA Delft, The Netherlands}

\author{Ophelia Crawford}
\affiliation{Riverlane, Cambridge, CB2 3BZ, UK} 
\author{Joonas Majaniemi}
\affiliation{Riverlane, Cambridge, CB2 3BZ, UK} 
\author{Marc Serra-Peralta}
\affiliation{QuTech and Delft Institute of Applied Mathematics, P.O. Box 5046, 2600 GA Delft, Delft University of Technology, The Netherlands}

\author{David Byfield}
\affiliation{Riverlane, Cambridge, CB2 3BZ, UK} 
\author{Boris Varbanov}
\affiliation{QuTech, Delft University of Technology, P.O. Box 5046, 2600 GA Delft, The Netherlands}

\author{Barbara M. Terhal}
\affiliation{QuTech and Delft Institute of Applied Mathematics, P.O. Box 5046, 2600 GA Delft, Delft University of Technology, The Netherlands}

\author{Leonardo DiCarlo}
\affiliation{QuTech and Kavli Institute of Nanoscience, Delft University of Technology, P.O. Box 5046, 2600 GA Delft, The Netherlands}

\author{Earl T. Campbell}
\affiliation{Riverlane, Cambridge, CB2 3BZ, UK} 
\affiliation{Department of Physics and Astronomy, University of Sheffield, UK}

\date{\today}
\setcounter{tocdepth}{2}

\begin{abstract}
Quantum error correction enables the preservation of logical qubits with a lower logical error rate than the physical error rate, with performance depending on the decoding method. Traditional decoding approaches rely on the binarization (`hardening') of readout data,  thereby ignoring valuable information embedded in the analog (`soft') readout signal. We present experimental results showcasing the advantages of incorporating soft information into the decoding process of a distance-three ($d=3$) bit-flip surface code with transmons. We encode each of the $16$ computational states that make up the logical state $\ket{0_{\mathrm{L}}}$, and protect them against bit-flip errors by performing repeated $Z$-basis stabilizer measurements. To infer the logical fidelity for the $\ket{0_{\mathrm{L}}}$ state, we average across the $16$ computational states and employ two decoding strategies: minimum-weight perfect matching and a recurrent neural network. Our results show a reduction of up to $6.8\%$ in the extracted logical error rate with the use of soft information. Decoding with soft information is widely applicable, independent of the physical qubit platform, and could allow for shorter readout durations, further minimizing logical error rates.
\end{abstract}
\maketitle

\begin{figure*}[t]
    \centering
    \includegraphics[width=0.9999\textwidth]{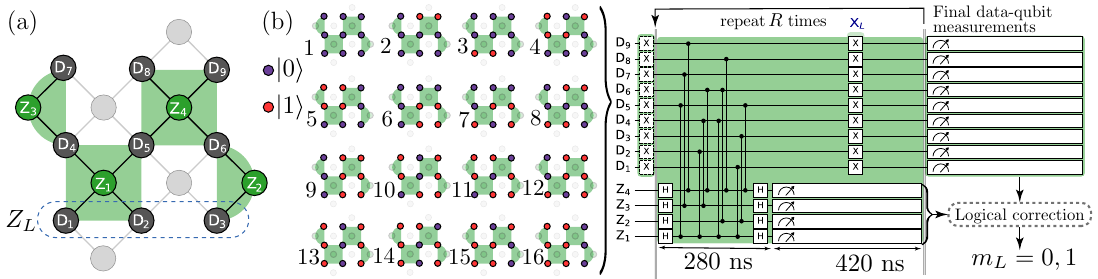}
    \caption{\textbf{Surface-13 QEC experiment}. (a) Device layout, with vertices indicating flux-tunable transmons, and edges denoting nearest-neighbor coupling via fixed-frequency resonators. Nine data qubits in a  $3\times3$ array  (labeled $D$, dark gray) are subject to 4 $Z$-basis parity checks realized using ancilla qubits (green). Light grey vertices and edges are not used. (b) Quantum circuit for experiment over $R$ QEC rounds, each round taking $700~\ns$. During ancilla measurement, a $X_{\mathrm{L}}$ operation implemented transversally with $\pi$ pulses on all data qubits is applied to average the logical error over the logical subspace. We show the 16 computational states that we average over. We employ various methods in post-processing to decode the measurements and determine the value of the logical observable, $m_{L}$~\cite{OBrien2017Density}.}
    \label{fig:Circuit}
\end{figure*}

\section{Introduction}

Small-scale quantum error correction (QEC) experiments have made significant progress over recent years, including fault-tolerant logical qubit initialization and measurement~\cite{ryan2021realization, Abobeih22}, correction of both bit- and phase-flip errors in a distance-three ($d=3$) code~\cite{Krinner22,Zhao22,Acharya22}, magic state distillation beyond break-even fidelity~\cite{Gupta24}, suppression of logical errors by scaling a surface code from $d = 3$ to $d = 5$~\cite{Acharya22}, and demonstration of logical gates~\cite{Bluvstein23}. The performance of these logical qubit experiments across various qubit platforms is dependent on the fidelity of physical quantum operations, the chosen QEC codes and circuits, and the decoders used to process QEC readout data. Common decoding approaches with access to analog information often rely on digitized (binary) qubit readout data as input to the decoder. The process of converting a continuous measurement signal to binary outcomes inevitably leads to a loss of information that reduces decoder performance.   

Pattison \textit{et al.}~\cite{pattison2021improved} proposed a method for incorporating this analog `soft' information in the decoding of QEC experiments, suggesting a potential $25\%$ improvement in the threshold compared to decoding with `hard' (binary) information. The advantage of using soft information has also been demonstrated on simulated data with neural-network (NN) decoders~\cite{varbanov2023neural, bausch2023learning}. Soft information decoding has also been realized for a single physical qubit measured via an ancilla in a spin-qubit system~\cite{Xue_2020} and for a superconducting-based QEC experiment with a simple error model assuming uniform qubit quality~\cite{sundaresan2023demonstrating}. The incorporation of soft information with variable qubit fidelity can in theory provide further benefit when decoding experimental data. However, this can be challenging as additional noise sources (e.g., leakage and other non-Markovian effects)  add complexity; therefore, the advantage of decoding with soft information is not guaranteed.

In this article, we demonstrate the use of soft information in the decoding of data obtained from a bit-flip $d=3$ code in a 17-qubit device using flux-tunable transmons with fixed coupling. Unlike a typical $d=3$ surface code, we repeatedly measure only $Z$-basis stabilizers, utilizing 13 out of the 17 qubits in the device [\cref{fig:Circuit}(a)]. This approach allows us to avoid problematic two-qubit gates between specific pairs of qubits that have strong interactions of the qubits with two-level system (TLS) defects~\cite{Muller19}. We refer to this experiment as Surface-13. We encode and stabilize each of the $16$ computational basis states, shown in \cref{fig:Circuit}(b), that are eigenstates of the $Z$-basis stabilizers and the logical operator, $ Z_{\mathrm{L}}$, with eigenvalues $+1$. We approximate the performance of the full logical state by averaging across these states. The code protects the logical state from single bit-flip errors, similar to the $d=3$ repetition code~\cite{Martinis15,Kelly15}. We employ two decoding strategies: a minimum-weight perfect matching (MWPM) decoder and a recurrent neural network (NN) decoder. For each strategy, we compare the performances of two variants: one with soft information and one without. With soft information, the extracted logical error rates are reduced by $6.8\%$ and $5\%$ for the MWPM and NN decoders, respectively.

\section{Experiment Configuration}

The experimental procedure begins by preparing the data-qubit register in one of the 16 physical computational states, as shown in \cref{fig:Circuit}(b). Next, repeated $Z$-basis stabilizer measurements are performed over a varying number of QEC rounds, $R$. Each round takes $700~\ns$, with $20~\ns$ and $60~\ns$ for single- and two-qubit gates, respectively, and $420~\ns$ for readout. The logical state is flipped during the ancilla measurement in each QEC round using the $X_{\mathrm{L}}=X^{\otimes 9}$ transversal gate to symmetrize the effect of relaxation ($T_1$) errors, minimizing the dependence on the input state. A final measurement of all data qubits is used to determine the observed logical $Z_{\mathrm{L}}$ outcome and compute final stabilizer measurements. The physical error rates of single-qubit gates, two-qubit gates, and readout are $0.1\%$, $1.6\%$, and $1.2\%$, respectively, averaging over the 13 qubits and 12 two-qubit gates used in the experiment. Further details about the device, calibration, and parity-check benchmarking are provided in \Cref{App:device_details,App:parity_check}.

The decoder determines whether the outcome $Z_{\mathrm{L}}$ needs to be corrected (flipped) based on the values of combinations of certain measurements (see below), and decoding success is declared if this corrected readout, $m_{L}$, matches the prepared state. We calculate $F_{\mathrm{L}}$, the logical fidelity, for a fixed $R\in\{ 1, 2,  4, 8, 16\}$ and each input state as the fraction of successfully decoded runs. Finally, $F_{\mathrm{L}}$ is averaged over the $16$ physical computational states, approximating the logical performance of the $\ket{0_{\mathrm{L}}}$ state.

Qubit readout is performed by probing the state-dependent transmission of a dedicated, dispersively-coupled readout-resonator mode to infer the qubit state, $\ket{j}$~\cite{Heinsoo18}. The readout pulse for each qubit has a rectangular envelope softened by a Gaussian filter of width $\sigma = 0.5~\ns$ (see \Cref{App:device_details} for additional details about readout calibration). After amplification~\cite{Bultink18}, the transmitted signal is down-converted to an intermediate frequency and the in-phase and in-quadrature components integrated over $420~\ns$ using optimal weight functions~\cite{Gambetta07,Ryan15,Magesan15}. The resulting two numbers, $I$ and $Q$, form the IQ signal $z=(I,Q)$ [\cref{fig:iq_gaussians_projected}(a)] comprising the soft information. The readout pulse envelope and signal integration are performed using  Z\"{u}rich Instruments UHFQA analyzers sampling at $1.8~\mathrm{GSa/s}$.

\section{Capturing Soft Information}
\label{Sec:Capturing_soft_info}

To transform an IQ signal $z \in \mathbb{R}^2$ to a binary measurement outcome $\hat{z}\in \{0, 1\}$, we apply a hardening map which we choose as the maximum-likelihood assignment. The hardened outcome is obtained by choosing $\hat{z} = 0$ if $P(0 \mid z)>P(1 \mid z)$ and $\hat{z} = 1$ otherwise, where $P(j \mid z)$ is the probability that the qubit was in state $\ket{j}$ just before the measurement given the observed IQ value $z$. Assuming the states $\ket{0}$ and $\ket{1}$ are equally likely, one has $P(j \mid z) \propto P(z \mid j)$. Therefore $\hat{z} = 1$ if $P(z \mid 1) > P(z \mid 0)$ and $\hat{z} = 0$ otherwise. When we consider the $\ket{2}$ state, we assign the hardened measurement outcome $j$ to correspond to the largest $P(z \mid j)$.

\begin{figure}[t]
    \centering
    \includegraphics[width=0.99\columnwidth]{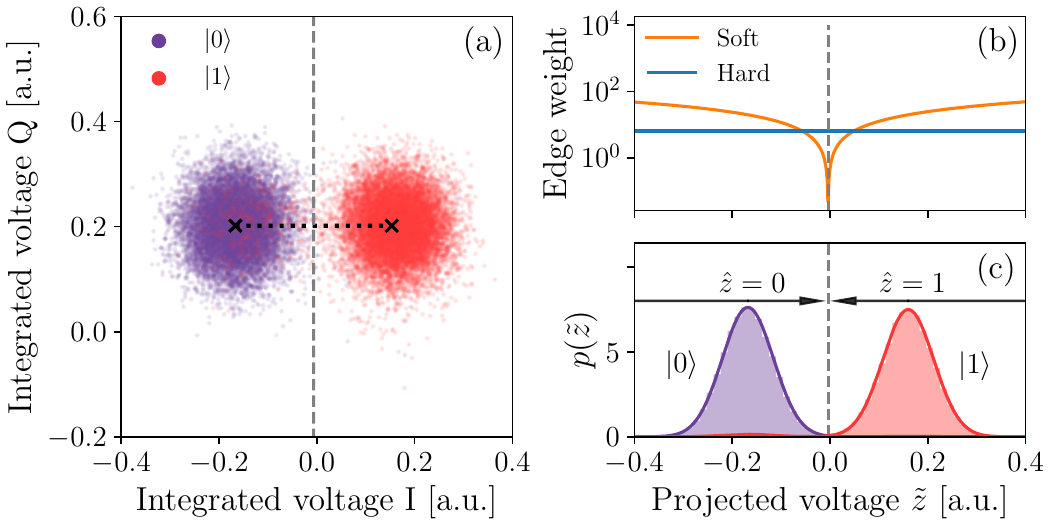}
    \caption{(a) The measurement response of the $\ket{0}$ and $\ket{1}$ states in IQ space for data qubit $D_6$, showing a projection line that connects the means of the two Gaussian peaks (black dotted line). (b) Edge weight as a function of projected voltage $\tilde{z}$ for soft and hard measurements; see \cref{eq:softedgeweight}. Measurement errors are most likely in the region $\tilde{z} \approx 0$ where the edge weight is minimized. (c) Histogram and fitted probability density function $P(\tilde{z}\mid j)$ for state preparations $j \in \{0, 1\}$.}
    \label{fig:iq_gaussians_projected}
\end{figure}

The probability density functions (PDFs) denoted $P(z \mid j)$ are combinations of two and three Gaussians when we do and do not consider the $\ket{2}$ state, respectively. We find that this heuristic model works well for both 2-state and 3-state discrimination. To determine the fit parameters of the Gaussian model, we use $1.3 \cdot 10^5$ calibration shots per state preparation $j \in \{0, 1, 2\}$ for each of the 13 transmons. To reduce the dimension of the problem in the case where we do not consider the $\ket{2}$ state, we project the IQ voltages to $\tilde{z} \in \mathbb{R}$ along the axis of symmetry [black dotted line in \cref{fig:iq_gaussians_projected}(a)]. We obtain the PDFs of the projected data $P(\tilde{z} \mid j)$ by decomposing $z = (\tilde{z}, z_\perp)$ to a parallel component $\tilde{z}$ and a perpendicular component $z_\perp$, giving $P(\tilde{z} \mid j)=\int P(\tilde{z}, z_\perp \mid j) dz_\perp$. Assuming the IQ responses of the computational basis measurements are symmetric along the axis joining the two centroids [marked by black crosses in \cref{fig:iq_gaussians_projected}(a)], the projection does not result in information loss. The hardened measurement outcomes are then obtained by comparing $P(\tilde{z} \mid 0)$ and $P(\tilde{z} \mid 1)$. Further information on our classification methods is given in \Cref{ap_sec:classifiers}.

\section{Minimum-Weight Perfect Matching Decoding}
\label{sec:MWPM}

In QEC experiments, detectors~\cite{higgott2023sparse} are selected combinations of binary measurement outcomes that have deterministic values in the absence of errors. A detector whose value has flipped from the error-free value is a defect. A decoder takes observed defects in a particular experiment and, using a model of the possible errors and the defects they result in, calculates the logical correction. In Surface-13, assuming circuit-level Pauli noise and with detectors defined as described below, each error results in at most two defects. As a result, it is possible to represent potential errors as edges in a graph -- the decoding graph -- with the nodes on either end representing the defects caused by the error (with a virtual node added for errors that only lead to a single defect). A matching decoder can thus be used, which matches pairs of observed defects along minimum~\cite{fowler_minimum_2014,dennis2002topological,higgott2021pymatching} or near-minimum~\cite{delfosse2021almost} weight paths within the graph, and thereby approximately finds the most probable errors that cause the observed defects. From the most probable errors, one can deduce whether a logical correction is necessary.

\begin{figure}[t]
    \centering
    \includegraphics[width=0.95\columnwidth]{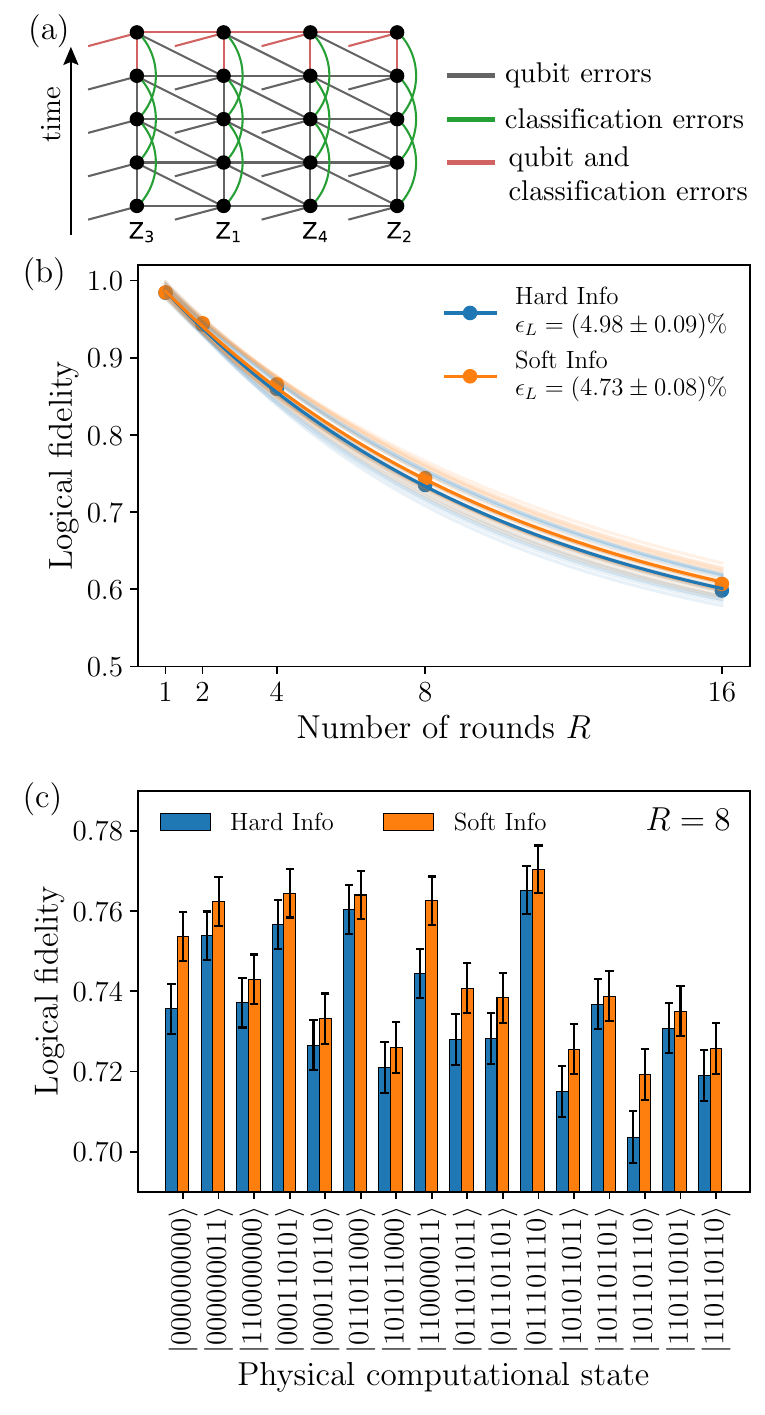}
    \caption{(a) Decoding graph showing different types of error mechanisms. The labels indicate the ancilla qubits associated with the detectors in each column. The soft MWPM decoder dynamically updates the weights of edges highlighted in green and red. (b) Logical fidelity of the MWPM decoder as a function of the number of rounds $R$, shown for each physical computational state that makes up $\ket{0_{\mathrm{L}}}$ (transparent curves) and averaged across all states (opaque curves). (c)~Logical fidelity at $R=8$, shown for each initial computational state as indexed in \Cref{fig:Circuit}b.}
    \label{fig:mwpmresults}
\end{figure}

Typically, QEC experiments are described assuming that ancilla qubits are reset following their measurement in every QEC round. However, this is not the case in our experiment. Nevertheless, without resetting qubits, suitable detectors can be chosen as $d_{i, r} = \hat{z}_{i, r} \oplus \hat{z}_{i, r-2}$
where $\hat{z}_{i, r} \in \{0, 1\}$ is the (hardened) measurement outcome of ancilla $i$ in round $r$ (see \Cref{App:no-reset} and ~\cite{varbanov2020leakage} for details).
We note that, in our case, the error-free detector values are always 0. A key difference with the mid-circuit reset case is that ancilla qubit errors (that change the qubit state) and measurement classification errors (where the inferred hardened measurement does not match the true qubit state) have different defect signatures, apart from in the final round (see \Cref{App:no-reset}). The structure of the decoding graph for the four-round experiment can be seen in \cref{fig:mwpmresults}(a).

To construct the decoding graph, one can define the probability of different error mechanisms and use a software tool such as Stim~\cite{gidney2021stim}. As we do not have direct knowledge of the noise, this graph may not accurately capture the true device noise. Therefore, we use a pairwise correlation method~\cite{spitz2018adaptive, chen2022calibrated, google2021exponential} to construct the graph for the MWPM decoder, whereby the decoding graph edge probabilities are inferred from the frequency of observed defects in the experimental data. In particular, this approach enables us to account for varying fidelities between different qubits. However, the pairwise correlation method is susceptible to numerical instabilities that we stabilize using a ``noise-floor graph"~\cite{google2021exponential} as described in \Cref{App:noise-floor}.  This is a crucial advance over previous experimental demonstrations of soft-information decoding~\cite{sundaresan2023demonstrating} where a graph derived from stim with uniform qubit fidelities was used. 

To use soft information with a MWPM decoder, we follow Ref.~\cite{pattison2021improved}. The edge corresponding to a measurement classification error is given a weight
\begin{equation}
    w = - \log \left [ \frac{P'(\tilde{z} \mid 1 - \hat{z}')}{P'(\tilde{z} \mid \hat{z}')} \right ],
\label{eq:softedgeweight}
\end{equation}
where $\hat{z}'$ is the inferred state after measurement. This is found by taking $\hat{z}' = 1$ if $P'(\tilde{z} \mid 1) > P'(\tilde{z} \mid 0)$ and 0 otherwise. The PDFs $P'(\tilde{z} \mid j')$ are obtained by keeping only the dominant Gaussian in the measurement PDFs -- this is to avoid including ancilla qubit errors during measurement in the classification error edge. Therefore, to incorporate soft information, we replace the weights calculated for the classification error edges using the pairwise correlation method with the weights from \cref{eq:softedgeweight}. This procedure is appropriate in all rounds but the final round of ancilla- and data-qubit measurements, where both qubit errors and classification errors have the same defect signature. In these cases, we instead (i) calculate the mean classification error for each measurement by averaging the per-shot errors; (ii) calculate the mean classification error for each edge from those for each measurement; (iii) remove the mean classification error from the edge probability; and (iv) include the per-shot classification error calculated from the soft readout information. Further information is given in \Cref{app:softinfo}.

\section{Neural Network Decoding}

Our second decoder -- the NN decoder -- can learn the noise model during training without making assumptions about it~\cite{bausch2023learning, varbanov2023neural, ueno_neo-qec_2022, baireuther2019neural}. NNs have flexible inputs that can include leakage or soft information, as well as non-uniform qubit fidelities. This again contrasts our work with previous soft-information decoding experiments that used a simpler noise model with uniform qubit fidelities~\cite{sundaresan2023demonstrating}.  
Recent work ~\cite{varbanov2023neural, bausch2023learning} has shown that NNs can achieve similar performance to computationally expensive (tensor network) decoders when evaluated on experimental data for the $d=3$ and $d=5$ surface code. Those networks were trained with simulated data, although Ref.~\cite{bausch2023learning} did a fine-tuning of their models with $\sim 2 \cdot 10^4$ experimental samples (while using $2 \cdot 10^9$ simulated samples for their main training). One may expect that the noise in the training and evaluation data should match to achieve the best performance.

\begin{figure}[!htb]
    \centering
    \includegraphics[width=0.95\columnwidth]{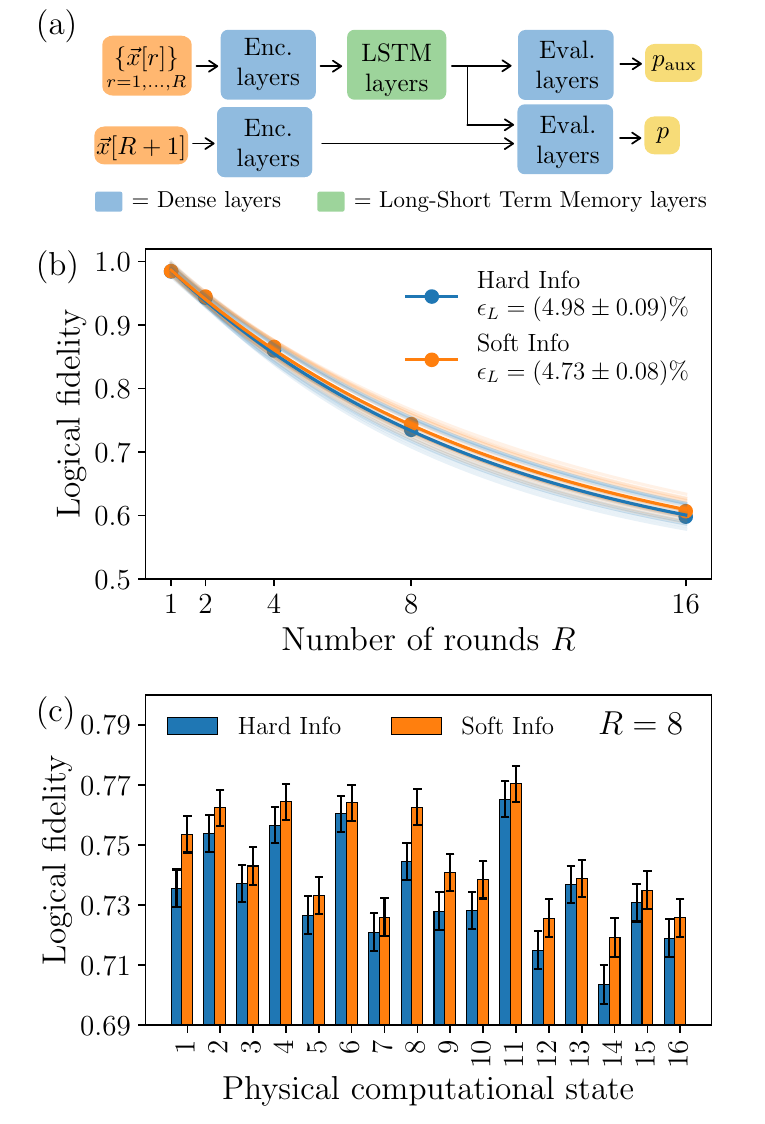}%{Figures/nn_combined_figure_v8bc.pdf}
    \caption{(a) Our NN architecture, a variant of Ref.~\cite{varbanov2023neural}. The input $\vec{x}[r]$ for the soft-NN contains the defect probability and leakage flag data of round $r$, while for the (hard-)NN it only contains the defect data. The vector $\vec{x}[R+1]$ contains the final-round defects.
    The output $p$ is the estimated probability that a logical error has happened and $p_{\mathrm{aux}}$ is only used to help the training (see \Cref{app_sec:nn_output}). (b) Logical fidelity of the NN decoder as a function of the number of rounds $R$, shown for each physical computational state that makes up $\ket{0_{\mathrm{L}}}$ (transparent curves) and averaged across all states (opaque curves). (c)~Logical fidelity at $R=8$, shown for each initial computational state as indexed in \Cref{fig:Circuit}b. 
    }
    \label{fig:architecture_nnresults}
\end{figure}

We train a NN decoder on experimental data and study the performance improvement when employing various components of the readout information available from the experiment. We use two architectures for our NN, corresponding to the network from Varbanov \textit{et al.}~\cite{varbanov2023neural}, and a variant of the network which includes encoding layers for handling different types of information [see \cref{fig:architecture_nnresults}(a)]. 
The inputs for our standard NN decoder are the observed defects. For our soft NN decoder, the inputs are the defect probabilities given the IQ values and the leakage flags, one for each ancilla qubit measurement. A leakage flag $l$ gives information about the qubit being in the computational space, i.e. $l =1$ if  $\hat{z} = 2$ and $l =0$ otherwise, where $\hat{z}$ is the hardened value of $z$ using the three-state classifier. For the final round when all data qubits are measured, we do not provide the decoder with any soft information to ensure that we do not make the task for the decoder deceptively simple~\cite{bausch2023learning}; this is a drawback of running the decoder only on the logical $\ket{0_{\mathrm{L}}}$ state and not on randomly chosen $\ket{0_{\mathrm{L}}}$ or $\ket{1_{\mathrm{L}}}$ (see the discussion in \Cref{app_sec:nn_data_leakage}). 

Due to the richer information of soft inputs, we can use larger networks than in the (hard-)NN case without encountering overfitting issues during training. The network performance when given different amounts of soft information is included in \Cref{app_sec:nn_output}, showing that the larger the amount, the better the logical performance. We follow the same training as Ref.~\cite{varbanov2023neural}, but with some different hyperparameters; and we use the ensembling technique from machine learning to improve the network performance without a time cost, but at a compute resource cost~\cite{naftaly1997optimal}.

\section{Logical Performance}

The extracted logical fidelity as a function of $R$ for the MWPM and NN decoders is shown in \cref{fig:mwpmresults,fig:architecture_nnresults}, respectively. The results are presented for the 16 physical computational states (transparent lines) and for their average (opaque lines), approximating the logical performance of the $\ket{0_{\mathrm{L}}}$ state. To find a logical error rate, $\epsilon_{\mathrm{L}}$, the decay of logical fidelity is fitted using the model
\begin{equation}\label{eq:logical_fidelity}
    \tilde{F}_{\mathrm{L}}(R) = \frac{1}{2}\left[ 1 + (1-2\epsilon_{\mathrm{L}})^{R-R_{0}}\right],
\end{equation}
where $\tilde{F}_{\mathrm{L}}$ indicates the fitted fidelity to the measured $F_{\mathrm{L}}$, and $R_{0}$ is a round offset parameter~\cite{OBrien2017Density}. For both decoders, including soft readout information enhances the logical performance, resulting in the reduction of $\epsilon_{\mathrm{L}}$  by $6.8\%$ and $5.0\%$ for MWPM and NN, respectively. We note that the error bars are different in the two cases due to the differing ways of splitting the dataset of $\sim 9\cdot 10^4$ samples per round and initial state. With the MWPM decoder, we use half the data to perform the pairwise correlation method to obtain the decoding graph, and half for obtaining the logical error probability. We then swap the data halves and average the logical fidelities, thereby using every shot to obtain the overall logical fidelity. With the NN decoder, $95\%$ of samples are used for training and validation and only $5\%$ of samples are used to estimate logical fidelities, leading to larger error bars.

\section{Summary}

We have experimentally demonstrated the benefits of using soft information in the decoding of a $d=3$ bit-flip surface code, utilizing $13$ qubits. We combined soft-information decoding with techniques that learn and account for variable qubit and gate fidelities, distinguishing our work from previous experiments~\cite{sundaresan2023demonstrating}. With soft information, the NN decoder achieves the best extracted logical error rate of $4.73\%$. It is crucial to note that, as we do not measure the $X$-basis stabilizers of the typical distance-$3$ surface code, the extracted $\epsilon_{\mathrm{L}}$ is likely underestimated (compared with $\epsilon_{\mathrm{L}}$ of $5.4\%$ for the standard $d=3$ surface code by Ref.~\cite{Krinner22} without leakage post-selection). However, the nature of the decoding problem will be the same and will benefit from the decoding optimizations explored in this paper. 

Despite the modest improvement in this work, simulations~\cite{pattison2021improved, bausch2023learning} suggest further advantages of soft information decoding: improvement in the error correction threshold, and increased suppression of logical errors as the code distance increases. Implementation of a leakage-aware decoder~\cite{suchara2015leakage} could also potentially enhance the logical performance with MWPM, but this has yet to be explored in experiment. With the NN decoder, it is unclear if the defect probabilities and leakage flags are the optimal way to present the information to the network, and this could be the subject of further investigation. Finally, our experiments utilized readout durations that were optimized for measurement fidelity. Calibrating the measurement duration for optimal logical performance~\cite{pattison2021improved} instead could potentially lead to higher logical performance.

\bigskip
\textit{Data and software availability}.- 
The source code of the NN decoder and the script to replicate the results are available in Ref.~\cite{Varbanov_qrennd_2023} and Ref.~\cite{Serra_surface13_nn_2024}, respectively. The script to analyze the experimental data can be found in Ref.~\cite{Serra_surface13_nn_2024}.

\section*{Acknowledgments}
This research is supported by the OpenSuperQPlus100 project (no.~101113946) of the EU Flagship on Quantum Technology (HORIZON-CL4-2022-QUANTUM-01-SGA), the Office of the Director of National Intelligence (ODNI), Intelligence Advanced Research Projects Activity (IARPA), via the U.S. Army Research Office Grant No. W911NF-16-1-0071, and QuTech NWO funding 2020-2024 -- Part I “Fundamental Research” with project number 601.QT.001-1. We acknowledge the use of the DelftBlue supercomputer~\cite{delftblue} for the training of the NNs. We thank G.~Calusine and W.~Oliver for providing the traveling-wave parametric amplifiers used in the readout amplification chain. H.A.'s contribution to this work was undertaken during his time at TU Delft.
The views and conclusions contained here are those of the authors and should not be interpreted as necessarily representing the official policies or endorsements, either expressed or implied, of the ODNI, IARPA, or the U.S. Government.

\bigskip
\textit{Author contributions}.- 
H.A. and J.F.M. calibrated the device and performed the experiment and data analysis. O.C., J.M. and D.B. performed and optimized the MWPM decoding. M.S.P. performed and optimized the NN decoding. E.T.C. and B.M.T. supervised the theory work. L.D.C. supervised the experimental work. H.A., O.C., M.S.P, J.M., E.T.C. wrote the manuscript with contributions from J.F.M., D.B., B.M.V., and B.M.T., and feedback from all coauthors.

\begin{figure*}[t!]
    \centering
    \includegraphics[width=0.90\textwidth]{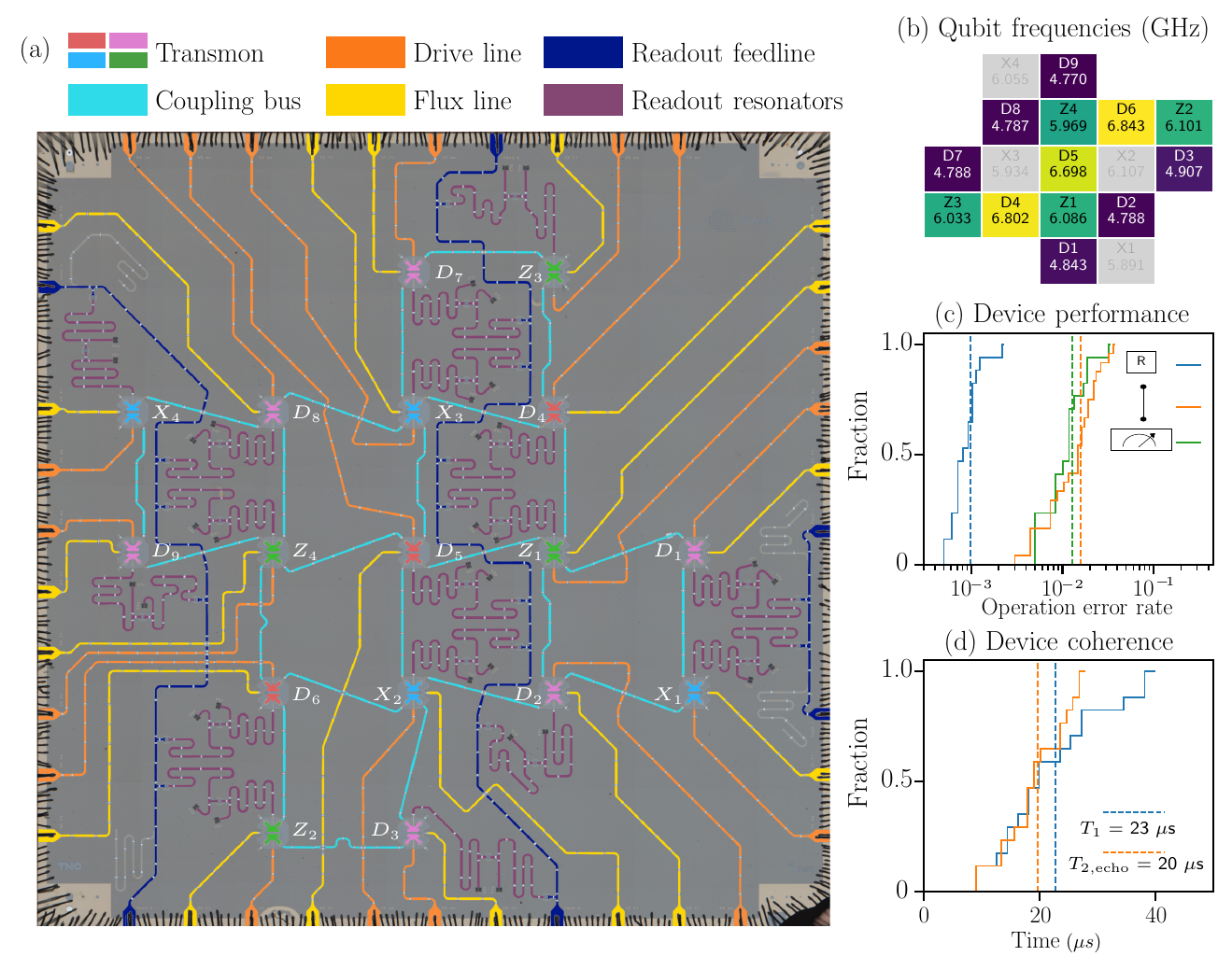}
    \caption{\textbf{Device characteristics}. (a) Optical image of the 17-transmon device, with added false color to emphasize different circuit elements. The device is connected to a printed circuit board using aluminum wirebonds, visible at the edges of the image. (b) Measured qubit transition frequencies with all transmons biased to their flux sweetspot. $X$-basis ancilla qubits (light gray) are not used in this experiment. (c) Cumulative distribution of error rates for single- and two-qubit gates, obtained by randomized benchmarking protocols with modifications to quantify leakage~\cite{Magesan12b,Wood18}, and average readout assignment fidelities, extracted from single-shot readout histograms~\cite{Walter17}. (d) Cumulative distribution of measured qubit relaxation time $T_{1}$ and echo dephasing time $T_{2, \mathrm{echo}}$. Dashed lines in (c) and (d) indicate the average over the 13 qubits used and the 12 two-qubit gates.}
    \label{fig:device_details}
\end{figure*}

\newpage

\appendix

\section{Device Overview}\label{App:device_details}

%device overview and overall performance
Our 17-transmon device~[\cref{fig:device_details}(a)] consists of a two-dimensional (2D) array of 9 data qubits and 8 ancilla qubits, designed for the distance-3 rotated surface code. Qubit transition frequencies are organized into three frequency groups: high-frequency qubits (red), mid-frequency qubits (blue/green), and low-frequency qubits (pink), as required for the pipelined QEC cycle proposed in Ref.~\cite{Versluis17}. 
Each transmon has a microwave drive line (orange) for single-qubit gates, a flux-control line (yellow) for two-qubit gates, and a dedicated pair of resonator modes (purple) distributed over three feedlines (blue) for fast dispersive readout with Purcell filtering~\cite{Walter17,Heinsoo18}. 
Nearest-neighbor transmons are coupled via dedicated coupling resonators (sky-blue)~\cite{Majer07}. Grounding airbridges (light gray) are fabricated across the device to interconnect the ground planes and to suppress unwanted modes of propagation. These airbridges are also added at the short-circuited end of each readout and Purcell resonator, allowing post-fabrication frequency trimming~\cite{Sanclemente23}. 
After biasing all transmons to their flux sweetspot, the measured qubit frequencies clearly exhibit three distinct frequency groups, as depicted in~\cref{fig:device_details}(b). These values are obtained from standard qubit spectroscopy. The average relaxation ($T_{1}$) and dephasing  ($T_{2, \mathrm{echo}}$) times of the 13 qubits used in the experiment are $23~\mu\mathrm{s}$ and $20~\mu\mathrm{s}$, respectively~[\cref{fig:device_details}d].
 
To counteract drift in optimal control parameters we automate re-calibration using dependency graphs~\cite{Kelly16}. The method, nicknamed graph-based tuneup (GBT)~\cite{AutoDepGraph17}, is based on Ref.~\cite{kelly18physical}. Single-qubit gates are autonomously calibrated with DRAG-type pulses to avoid phase error and to suppress leakage~\cite{Motzoi09,Chow10b}, and benchmarked using single-qubit randomized benchmarking protocols~\cite{Magesan11}. The average error of the calibrated single-qubit gates [\cref{fig:device_details}(c)] across $13$ qubits reaches $0.1\%$ with a leakage rate of $10^{-4}$. All single-qubit gates have $20~\ns$ duration.

Two-qubit controlled-$Z$ (CZ) gates are realized using sudden Net-Zero flux pulses~\cite{Negirneac20}. The QEC cycle in this experiment requires 12 CZ gates executed in 4 steps, each step performing 3 CZ gates in parallel. This introduces new constraints compared to tuning an individual CZ gates. For instance, parallel CZ gates must be temporally aligned to avoid overlapping with unwanted interaction zones on the way to, from, or at the intended avoided crossings. Moreover, simultaneous operations in time (vertical) and space (horizontal) may induce extra errors due to various crosstalk effects, such as residual $ZZ$ coupling, microwave cross-driving, and flux crosstalk. To address these non-trivial errors, we introduce two main calibration strategies into a GBT procedure: vertical and horizontal calibrations (VC and HC). These tune simultaneous CZ gates in time and space as block units~\cite{Ali_aps_22}. This approach absorbs some of the flux and residual-$ZZ$ crosstalk errors. After calibration, GBT benchmarks the calibrated gates with two-qubit interleaved randomized benchmarking protocols with leakage modification~\cite{Magesan12b,Wood18}. The individual benchmarking of the $12$ CZ gate reveals an average error of $1.6\%$ with a $0.24\%$ leakage. All CZ gates have $60~\ns$ duration.

Readout calibration is performed in three main steps, realized manually and not with a GBT procedure. In the first, readout spectroscopy is performed at fixed pulse duration ($200~\ns$-$300~\ns$) to identify the optimal frequency maximizing the distance between the two complex transmission vectors $S_{21}^{|0\rangle}$ and $S_{21}^{|1\rangle}$ in the IQ plane. The second step involves a 2D optimization over pulse frequency and amplitude. The goal is to determine readout pulse parameters that minimize a weighted combination of readout assignment error ($\varepsilon_\mathrm{RO}$), and measurement quantum non-demolition (QND) probabilities ($P_{\mathrm{QND}}$ and $P_{\mathrm{QND}\pi}$). These probabilities are obtained using the method of Ref.~\cite{Chen22}. The final step verifies if photons are fully depleted from the resonator within the target total readout time, $420~\ns$, using an ALLXY gate sequence between two measurements~\cite{Bultink16}. By comparing the ALLXY pattern obtained to the ideal staircase, we can determine if the time dedicated for photon depletion is sufficient to not affect follow-up gate operations. 

After calibrating optimal readout integration weights~\cite{Bultink16}, we proceed to benchmark various readout metrics such as $\varepsilon_\mathrm{RO}$ and standard readout QND ($F_\mathrm{QND}$) using the measurement butterfly technique~\cite{Marques23}. The average $\varepsilon_\mathrm{RO}$ [\cref{fig:device_details}(c)] is $1.2\%$, extracted from the single-shot histograms. We also perform simultaneous multiplexed readout of all 13 qubits, constructing assignment probability and cross-fidelity matrices~\cite{Heinsoo18}. The average multiplexed readout error rate is $1.6\%$, indicating that readout crosstalk is small. Moreover, the average $F_\mathrm{QND}$ for the four $Z$-basis ancillas is $95.3\%$ considering a three-level transmon~\cite{Marques23}. This also yields an average leakage rate due to ancilla measurement of $0.14\%$, predominantly from $\ket{1}$.

\setcounter{lastequationbeforesection}{\number\value{equation}}
\section{Benchmarking of Parity Checks}\label{App:parity_check}
\setcounter{equation}{\number\value{lastequationbeforesection}}

\begin{figure*}[t]
    \centering
    \includegraphics[width=0.9\textwidth]{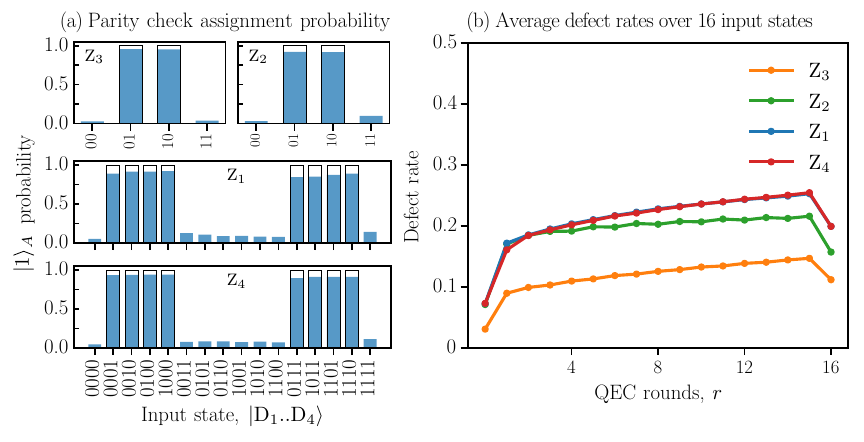}
    \caption{\textbf{Parity check benchmarking}. (a) Benchmarking of the assignment fidelity for four stabilizer measurements: $Z_{D_{4}}Z_{D_{7}}$, $Z_{D_{6}}Z_{D_{3}}$, $Z_{D_{4}}Z_{D_{5}}Z_{D_{2}}Z_{D_{1}}$, and $Z_{D_{9}}Z_{D_{9}}Z_{D_{5}}Z_{D_{6}}$. (b) The average defect rate as a function of QEC rounds for each of the four $Z$-basis stabilizers across the 16 input states.}
    \label{fig:PC_defect_s13}
\end{figure*}

With the individual building blocks calibrated, we proceed to calibrate the four $Z$-basis stabilizer measurements as parallel block units using VC and HC strategies, as discussed  in~\Cref{App:device_details}. The average probabilities of correctly assigning the parity operator $\Pi_{i}Z_i$ are measured as a function of the input computational states of the data-qubit register. The measured probabilities (solid blue bars)~[\cref{fig:PC_defect_s13}(a)] are compared with the ideal ones (black wireframe) to obtain average parity assignment fidelities, $96.3\%$, $92.8\%$, $89.9\%$, and $92.3\%$ for $\mathrm{Z}_3$, $\mathrm{Z}_2$, $\mathrm{Z}_1$, $\mathrm{Z}_4$, respectively. These results are obtained after mitigating residual excitation effects by post-selection on a pre-measurement~\cite{Marques23}.

The defect rates (see \Cref{Sec:Capturing_soft_info} for the definition of a defect), reflect the incidence of physical qubit errors (bit-flip and readout errors) detected throughout the rounds $r$, where $r \in\{1,2,...~R\}$. For each of the four $Z$-basis stabilizers, the defect rate~[\cref{fig:PC_defect_s13}(b)] is presented over 16 QEC rounds and averaged across the 16 physical computational input states. The sharp increase in the defect rate between rounds $r=1$ and $r=2$ is due to the low initialization error rates and the detection of errors occurring during the ancilla-qubit measurements in the first round. At the boundary round $r=16$, the defects are obtained using the final data-qubit measurements, which, given the low readout error rates, lead to the observed decrease in the defect rate. Over rounds, defect rates gradually build up until levelling at approximately $15\%$, $22\%$, $25\%$, and $25\%$ for $\mathrm{Z}_3$, $\mathrm{Z}_2$, $\mathrm{Z}_1$, $\mathrm{Z}_4$, respectively. The build-up may be due to leakage~\cite{Chen21,Krinner22,Acharya22,Marques23}.

\setcounter{lastequationbeforesection}{\number\value{equation}}
\section{Decoding Graph}
\setcounter{equation}{\number\value{lastequationbeforesection}}
\label{App:no-reset}
In this Appendix, we describe in further detail the effect of not using mid-circuit resets on the decoding graph. Here, we consider specifically Surface-13, but similar ideas apply to the full surface code and other stabilizer codes.

As discussed in \Cref{sec:MWPM}, in order to detect errors in stabilizer codes, it is typical to define detectors, combinations of binary measurement outcomes that have deterministic values in the absence of errors. We refer to detectors whose value has flipped from the expected error-free value as defects. A decoder takes observed defects in a particular experiment and, using a model of the possible errors and the defects they result in, predicts how the logical state has been affected by the errors. It is desirable for errors to result in a maximum of two defects, as this enables a matching decoder to be used, which can efficiently find the most probable~\cite{fowler_minimum_2014,dennis2002topological,higgott2021pymatching} errors that cause the observed defects.

In Surface-13, the stabilizers are, in the bulk, weight-four $Z$ operators. On the boundaries, the stabilizers are of lower weight. In general, each data qubit is involved in two $Z$-type stabilizers, or one on the boundary. The experiment proceeds by using ancilla qubits to measure the stabilizers for some number of rounds, $R$, with $s_{i, r}$ the $i$th stabilizer outcome in the $r$th round. An $X$ error on a data qubit in round $r$ will change the value of the outcomes of the $Z$ stabilizers involving that data qubit from round $r$ onwards. We define the detectors $d_{i, r}$ to be the difference between stabilizer measurements in adjacent rounds, so that
\begin{equation}
    d_{i, r} = s_{i, r} \oplus s_{i, r-1}. \label{eq:detector_1}
\end{equation}
We begin Surface-13 with initial data qubit states that are eigenstates of the $Z$-type stabilizers with eigenvalues +1; therefore, we set $s_{i, 0} = 0$. At the end of the experiment, the data qubits are measured in the Z-basis; these measurements can be used to construct $s_{i, R+1}$ outcomes for the stabilizers.

With the definition in \cref{eq:detector_1}, an $X$ data qubit error before a QEC round results in a maximum of two defects, as desired. We note that, in general, there may be multiple errors that result in the same defect signature. The probabilities of these errors are typically combined to give a single edge weight.

We now consider the effect of ancilla qubit and measurement classification errors. An ancilla qubit error in round $r$ on the qubit used to measure the $i$th stabilizer will change the stabilizer measurement outcome in the $r$th round, thus resulting in defects $d_{i, r}$ and $d_{i, r+1}$. The effect of classification errors depends on how the stabilizer outcomes are obtained. If the ancilla qubits are reset after measurement, $s_{i, r} = \hat{z}_{i, r}$, where $\hat{z}_{i, r}$ is the hardened measurement outcome on the $i$th ancilla at round $r$, and therefore a classification error in round $r$ results in the same defects as an ancilla qubit error in round $r$. However, if the ancilla qubits are not reset after measurement, as is the case in this paper, $s_{i, r} = \hat{z}_{i, r} \oplus \hat{z}_{i, r-1}$. Therefore, with respect to measurements, the detectors are defined as
\begin{equation}
    d_{i, r} = \hat{z}_{i, r} \oplus \hat{z}_{i, r-2}, \label{eq:detector_2}
\end{equation}
for $2 \leq r  \leq R$. In this case, a classification error in measurement $\hat{z}_{i, r}$ causes defects $d_{i, r}$ and $d_{i, r+2}$. In both cases, a data qubit measurement classification error looks like a data qubit Pauli error between rounds $R$ and $R+1$. We also define $d_{i, 1} = \hat{z}_{i,1}$.

In this error model, we assume that two independent events are possible during measurement -- an ancilla qubit error and a classification error. In practice, these are not independent events as both are affected by $T_{1}$ processes. However, should an error occur that causes the qubit to decay from the $\ket{1}$ to the $\ket{0}$ state during measurement, it can either be viewed as an ancilla qubit error before measurement (if the inferred hardened measurement outcome is 0) or an ancilla qubit error after measurement (if the inferred hardened measurement outcome is 1). Therefore, these coupled events can be viewed as a single ancilla qubit error. We note, however, that such qubit errors are not symmetric, and thus using a single edge weight is an approximation.

We note that we have not discussed mid-round qubit errors that result in so-called hook errors, as we have focused on explaining the differences between the decoding graphs with and without mid-circuit reset. The hook errors in the two cases will be identical.

\subsection{Noise-floor Graph}
\label{App:noise-floor}
 As discussed in the main text, the pairwise correlation method can be subject to numerical instabilities. We thus use a ``noise-floor graph" that has specific values for each edge in the decoding graph. These instabilities can arise due to the finite data and the approximation that there are no error mechanisms resulting in more than two defects, such as leakage. The impact is more pronounced at the boundary of the code lattice where single defects occur~\cite{varbanovPHD}.
 
The lower-bound error parameters used to construct the noise-floor decoding graph are given in \Cref{ap_tab:noise_floor}. The operation times are taken to be the same as the real device. We note, however, that the other parameters are not the same as those stated in \Cref{App:device_details}. This is because the parameters here set a lower bound on the error probabilities and should only be used when the pairwise correlation method gives unfeasibly low values. Whilst extensive exploration of the parameters was not undertaken, the ones stated here were found to give good performance and several other options did not result in significant changes to the results.

Each probability, $p$, is the probability of a depolarising error after the specified operation. These are defined so that, for single-qubit gates, resets and measurements, the probability of applying the Pauli error $W$ is given by
\begin{equation}
    p_{W} = \frac{p}{3},
\end{equation}
for $W \in \{ X,\, Y,\, Z \}$. For two-qubit gates, the probability of applying the Pauli $VW$ is given by
\begin{equation}
    p_{VW} = \frac{p}{15},
\end{equation}
where $V,\, W \in \{I,\, X,\, Y,\, Z \}$, excluding $V = W = I$. Idle noise is incorporated by Pauli twirling the amplitude damping and dephasing channel to give, for an idling duration $t$, ~\cite{sarvepalli2009asymmetric}
\begin{align}
    p_{X}(t) &= p_{Y}(t) = \frac{1}{4} \left(1 - e^{-t/T_1} \right), \\
    p_Z(t) &= \frac{1}{2} \left(1 - e^{-t/T_2} \right) - \frac{1}{4} \left(1 - e^{-t/T_1} \right).
\end{align}
We note that, as $T_{1}$ = $T_{2}$ in our model, $p_{X}(t) = p_{Y}(t) = p_{Z}(t)$.
The noise-floor graph is derived from the circuit containing the appropriate parameters using Stim~\cite{gidney2021stim}.
\begin{table}[h]
    \centering
    \caption{Lower-bound noise parameters used in the experimental graph derivation. These parameters are used to fix the minimum values of each edge.}
    \begin{tabular}{|c|c|}
         \hline
    Parameter & Value\\
    \hline
    single-qubit gate error probability & $0.5 \cdot 10^{-3}$ \\
    two-qubit gate error probability & $5 \cdot 10^{-3}$ \\
    reset error probability & $0.0$ \\
    measurement qubit error probability & $1 \cdot 10^{-3}$ \\
    measurement classification error probability & $1 \cdot 10^{-3}$ \\
    $T_{1}$ & $30~\mu$s  \\
    $T_{2}$ & $30~\mu$s  \\
    single-qubit gate time & $20~\ns$ \\
    two-qubit gate time & $60~\ns$ \\
    measurement time & $420~\ns$ \\
    \hline
    \end{tabular}
    \label{ap_tab:noise_floor}
\end{table}

\setcounter{lastequationbeforesection}{\number\value{equation}}
\section{Details of the Neural Network Decoder}
\setcounter{equation}{\number\value{lastequationbeforesection}}
\label{app_sec:nn}

\subsection{NN inputs, outputs and decoding success} \label{app_sec:nn_output}

The inputs provided to the NN decoder consist of the defects, the defect probabilities, and the leakage flags [see \Cref{fig:architecture_nnresults}(a)]. The only elements not described in the main text are the defect probabilities. These are obtained following Ref.~\cite{varbanov2023neural} and using the two-state readout classifier from the main text and \Cref{ap_sec:gaussian_mixture_model}. First, we express the probability of the measured qubit ‘having been in the state’ $\ket{j}$ ($j \in \{0, 1\}$) given the IQ value $z$ as
\begin{equation} \label{app_eq:prob_k_z}
    P(j|z) = \frac{P(z|j)P(j)}{\sum_{i}P(z|i)P(i)},
\end{equation}
with $P(j)$ the probability that the qubit was in state $\ket{j}$. 
We define the \textit{incoming defects} in the bulk as $\tilde{d}_{i, r} = k_{i, r} \oplus k_{i, r-2}$, where $k_{i, r}$ is the state of the ancilla $i$ before its measurement in round $r$. 

Although we do not have access to the incoming defects, we can estimate the probability that the defect $\tilde{d}_{i, r}$ has been triggered given $z_{i, r}$ and $z_{i, r-2}$ (named \textit{defect probability}) by
\begin{align}\label{app_eq:d_prob} \nonumber
    & P(\tilde{d}_{i, r} = 1 | z_{i, r}, z_{i, r-2}) =  \\ \nonumber & P(k_{i,r}=0|z_{i,r}) P(k_{i,r-2}=1|z_{i,r-2}) \\ & + P(k_{i,r}=1|z_{i,r}) P(k_{i,r-2}=0|z_{i,r-2}).
\end{align}
As we can incorrectly infer $k_{i,r}$ from $z_{i,r}$, the defect probabilities include assignment errors. Note that digitizing $P(\tilde{d}_{i, r} = 1 | z_{i, r}, z_{i, r-2})$ leads to the ``standard" defects defined in the main text~\cite{bausch2023learning}. 
Using defect probabilities allows us to infer the defect reliability, e.g. $P(k_{i,r}=0|z_{i,r}) \approx P(k_{i,r}=1|z_{i,r})$ leads to $P(d_{i, r} = 1 | z_{i, r}, z_{i, r-2}) \approx 1/2$, which is in-between 0 and 1, thus uncertain. 

We have used the incoming defects and not the ``standard" defects for deriving the defect probabilities because $P(\hat{z}|z)$ is always $0$ or $1$ as $\hat{z}$ is completely determined by $z$. In that case, we would not have any soft information and the defect probabilities would correspond to the defects. Remember that we do not use the defect probabilities of the final round as explained in \Cref{app_sec:nn_data_leakage} and that we assume $P(0) = P(1) = 1/2$ when calculating $P(d_{i, r} = 1 | z_{i, r}, z_{i, r-2})$. 

\begin{figure}[htb]
\includegraphics[width=0.4\textwidth]{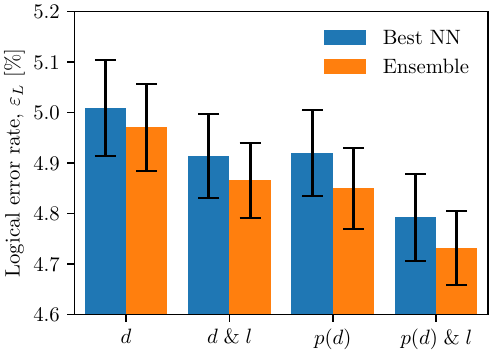}
\caption{Logical error rates of the NN decoders when given different inputs: the label $d$ corresponds to defects, $l$ to leakage flags, and the label $p(d)$ corresponds to defect probabilities given the soft information. ``Best NN" (blue) and ``Ensemble" (orange) correspond to the architectures with the lowest logical error rate for a single NN and for an ensemble of 5 NNs, respectively, as described in Section \ref{app_sec:ensembling}.}
\label{fig:nn_inputs}
\end{figure}

For completeness, we study the performance of the NNs given four combinations of inputs: (a) defects, (b) defects and leakage flags, (c) defect probabilities, and (d) defect probabilities and leakage flags. The results in \Cref{fig:nn_inputs} show that the networks can process the richer inputs to improve their performance. The reason for not giving the network the soft measurement outcomes $z$ as input directly (but the defect probabilities instead) is that we have found that the NN decoder does not perform well on the soft measurements~\cite{varbanov2023neural}; effectively, the NN has to additionally learn the defects, which is possible with larger NNs as in Ref.~\cite{bausch2023learning}. 

The NN gives two outputs, $p$ and $p_{\mathrm{aux}}$, which correspond to the estimated probabilities that a logical flip has occurred during the given sample. The output $p$ has been calculated using all the information given to the NN, while $p_{\mathrm{aux}}$ does not use the final round data; see \Cref{fig:architecture_nnresults}(a). We train the NN based on its accuracy in both $p$ and $p_{\mathrm{aux}}$ because the latter helps the NN to not focus only on the final round but decode based on the full QEC data~\cite{baireuther2018machine, varbanov2023neural}. Note that the outputs correspond to physical probabilities (i.e., $p, p_{\mathrm{aux}} \in [0,1]$) thanks to using sigmoid activation functions in the last layer of the NN architecture. 

To determine whether the neural network decoded the QEC data correctly or not, the output $p$ is used as follows. If $p\geq 1/2$, we set the logical flip bit to $b=1$; if $p< 1/2$, we set the logical flip bit to $b=0$. On the basis of the final data qubit measurements, we compute the (uncorrected) logical $Z_{\mathrm{L}}$ as a bit $z_{\rm out} \in \{0,1\}$. We take the logical input state $z_{\rm in}$ (in our experiments $z_{\rm in}=0$ always as we prepare $\ket{0_{\mathrm{L}}}$ in all cases) and mark the run as successful when $z_{\rm in}\oplus z_{\rm out} \oplus b=0$, and unsuccessful when $z_{\rm in}\oplus z_{\rm out} \oplus b=1$.

\subsection{Learning the final logical measurement} \label{app_sec:nn_data_leakage}

In machine learning, one needs to be careful about the information given to the network. For example, the neural network could predict the logical bit flip correction $b$ without using the defect information gathered over multiple rounds. In particular, given that in our experiment we only start with state $\ket{0_{\mathrm{L}}}$, the neural network could just output the value of $b$ such that $b \oplus z_{\rm out}=0$. It can do this by learning $z_{\rm out}$, and hence the network should get no explicit information about $z_{\rm out}$. 
While it is true that we do not directly provide the data-qubit measurement outcomes to the NN decoder, there still might be some partial information provided by the defect probabilities or the leakage flags, e.g. when a transmon in state $\ket{2}$ is more likely declared as a $\ket{1}$ than a $\ket{0}$; see \cref{ap_fig:2D_IQ_ancillas} for transmon $Z_2$. Note that this issue does not occur if one randomly trains and validates with either $\ket{0_{\mathrm{L}}}$ or $\ket{1_{\mathrm{L}}}$ since then the optimal $b=z_{\rm in}\oplus z_{\rm out}$ and $z_{\rm in}$ is a random bit unknown to the network.

We indeed observe an abnormally high performance for a single network when using soft information in the final round, with $\epsilon_{\mathrm{L}} \sim 4.2\%$ when using defect probabilities and $\epsilon_{\mathrm{L}} \sim 4.1\%$ if we include the leakage flags too. Such an increase suggests that the NN can partially infer $z_{\rm out}$ and thus `knows' how to set $b$. 
This phenomenon does not occur when giving only the leakage flags of data qubit measurements, which instead leads to $\epsilon_{\mathrm{L}} \sim 4.9\%$. Nevertheless, due to the reasons explained above, we decided to not include the leakage flags in the final round to be sure we do not provide any information about $z_{\rm out}$ to the NN. 

We note that Ref.~\cite{bausch2023learning} also opted to just give the defects in the final round because of the possibility that the NN decoder can infer the measured logical outcome from the final-round soft information.

\subsection{Ensembling} \label{app_sec:ensembling}

Ensembling is a machine-learning technique to improve the network performance without costing more time, as networks can be trained and evaluated in parallel~\cite{naftaly1997optimal}.
It consists of averaging the outputs, $\{p_i\}$, of a set of NNs, to obtain a single more accurate prediction, $\tilde{p}$. One can think that the improvement is due to ``averaging out" the errors in the models~\cite{naftaly1997optimal}. 
Reference~\cite{bausch2023learning} trained 20 networks with different random seeds and averaged their outputs with the geometric mean. 
In this work, the output `average', $\tilde{p}$, is given by
\begin{equation}
\log \left( \frac{1-\tilde{p}}{\tilde{p}} \right) = \frac{1}{5} \sum_{i=1}^5 \log \left( \frac{1 - p_i}{p_i} \right),
\end{equation}
with $\{p_i\}$ the predictions of 5 individual NNs of the logical flip probability; see \Cref{app_sec:nn_output}. 
This expression follows the approach from the repeated qubit readout with soft information~\cite{dAnjou2021generalized}, which is optimal if the values are independently sampled from the same distribution. Once $\tilde{p}$ is determined, we threshold it to set the flip bit $b$ as described in Section \ref{app_sec:nn_output}.

\subsection{Network sizes, training hyperparameters and dataset} \label{app_sec:nn_sizes_hyperparameters}

\begin{figure*}[htb]
\includegraphics{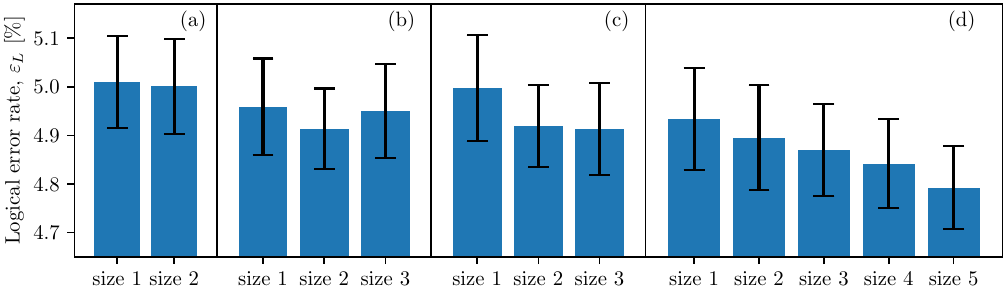}
\caption{Logical performance of the NNs given the four input combinations and different network sizes. The NN input $\vec{x}[r]$ in panel (a) are the defects, in (b) the defects and leakage flags, in (c) the defect probabilities, and in (d) the defect probabilities and the leakage flags. The NN sizes are summarized in \Cref{tab:nn_summary}. }
\label{fig:nn_architectures}
\end{figure*}

Due to the different amounts of information in each input, the NNs in \Cref{fig:architecture_nnresults} and \Cref{fig:nn_inputs} have different sizes to maximize their performance without encountering overfitting issues. In \cref{fig:nn_architectures}, the size of the network is increased given a set of inputs until overfitting degraded the performance or there was no further improvement. 
The specific sizes and hyperparameters of the NNs shown in the figures are summarized in \Cref{tab:nn_summary}. These hyperparameters are the same as in Ref.~\cite{varbanov2023neural} but with the following changes: (1) reducing the batch size to avoid overfitting, as the experimental dataset is smaller, and (2) decreasing the learning rate for the large NNs.
For comparison, the NN in Ref.~\cite{bausch2023learning} for a $d=3$ surface code uses 5.4 million free parameters, a learning rate of $3.5 \cdot 10^{-4}$, and a batch size of 256. 
The number of free parameters in a NN is related to its capacity to learn and generalize from data, the learning rate is related to the step size at which NN parameters are optimized, and the batch size is the number of training samples used in a single iteration of gradient descent. 

\begin{table*}[htb]
\centering
\caption{NN sizes and the training hyperparameters used in this work. $n_i$ refers to the number of layers in block $i \in \{\text{Enc}, \text{LSTM}, \text{Eval}\}$ and $d_i$ the dimension of these layers. If $n_{\text{Enc}}$ is not specified, the network does not have Encoding layers. The blocks are shown in Fig 4(a). The number of free parameters depends on the given input combination, but the changes are of the order of $\sim$5k. }
\label{tab:nn_summary}
\begin{tabular}{c|ccccccc|ccc} \hline
Label & $n_{\text{Enc}}$ & $d_{\text{Enc}}$ & $n_{\text{LSTM}}$ & $d_{\text{LSTM}}$ & $n_{\text{Eval}}$ & $d_{\text{Eval}}$ & \begin{tabular}[c]{@{}c@{}}\# free \\ parameters\end{tabular} & \begin{tabular}[c]{@{}c@{}}batch \\ size\end{tabular} & \begin{tabular}[c]{@{}c@{}}learning \\ rate\end{tabular} & \begin{tabular}[c]{@{}c@{}}dropout \\ rate\end{tabular} \\ \hline
size 1 &  &  & 2 & 90 & 2 & 90 & $\sim$ 115k & 64 & $5\cdot 10^{-4}$ & 20\% \\
size 2 & 2 & 32 & 2 & 100 & 2 & 100 & $\sim$ 160k & 64 & $2\cdot 10^{-4}$ & 20\% \\
size 3 & 2 & 64 & 2 & 120 & 2 & 120 & $\sim$ 250k & 64 & $2\cdot 10^{-4}$ & 22\% \\
size 4 & 2 & 90 & 3 & 100 & 2 & 100 & $\sim$ 285k & 64 & $2\cdot 10^{-4}$ & 22\% \\
size 5 & 2 & 100 & 3 & 100 & 2 & 100 & $\sim$ 290k & 64 & $2\cdot 10^{-4}$ & 20\% \\ \hline
\end{tabular}
\end{table*}

\begin{figure}[htb]
\includegraphics{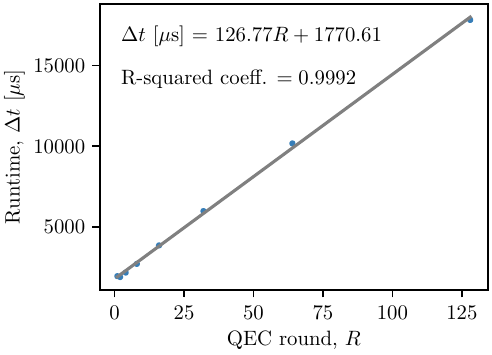}
\caption{Evaluation runtime for the size-5 network with batch size = 1. The line corresponds to a linear regression where the R-squared coefficient shows that the fit is appropriate. The inputs for the NN in this calculation are created at random (not based on experimental data) because we are not interested in the logical performance and the number of operations the NN needs to perform only depends on the number of rounds. In particular, the number of operations in the LSTM and encoding layers grows linearly with $R$, but for the evaluation layers it is constant. Therefore, we can associate the $y$ intercept as the time required for the evaluation layers and the slope as the time spent on the encoding and LSTM layers. 
Each point is the average of $5\cdot 10^4$ samples. }
\label{ap_fig:runtime}
\end{figure}

The splitting of the experimental dataset into the three sets of \textit{training}, \textit{validation}, and \textit{testing} was done as follows. For each initial state and each number of rounds, (1) randomly pick $5\cdot 10^3$ samples from the given data and store them in the \textit{testing} dataset, (2) randomly select 90\% of the remaining samples and store them in the \textit{training} dataset, and (3) store the rest in the \textit{validation} dataset. The reason for this choice is to ensure that the datasets are not accidentally biased towards an initial state or number of rounds. 
After the splitting, we have a training dataset consisting of $6.9\cdot 10^6$ samples, a validation dataset with $7.6\cdot 10^5$, and a testing dataset with $4\cdot 10^5$. 
Note that the NN has been trained and tested on the same number of rounds because the experiment only goes until $R=16$. However, in longer memory experiments, the NN should be trained only up to a ``low" number of rounds to avoid long trainings. 

The training for each single NN was carried out on an NVIDIA Tesla V100S GPU and lasted around 10~hours for the smallest size and 23~hours for the largest one when using the training dataset consisting of $6.9 \cdot 10^6$ samples with $7.6 \cdot 10^5$ samples for validation. 
The evaluation of the network performance was done on an Intel Core(TM) i7-8650U CPU @ 1.90GHz $\times$ 4. We estimate that it takes $\sim 127 \;\mu\text{s}$ per QEC cycle and $\sim 1.77\;\text{ms}$ for the final round when running a size-5 NN with soft-information; see \cref{ap_fig:runtime}. The same NN without soft information takes $\sim 124 \;\mu\text{s}$ per QEC cycle and $\sim 1.57\;\text{ms}$ for the final round. The size-2 NN used in the main text to decode data without soft information takes $\sim 82 \;\mu\text{s}$ per QEC cycle and $\sim 1.35\;\text{ms}$ for the final round. 
For comparison, the NN in Ref.~\cite{bausch2023learning} for a $d=3$ surface code takes $\sim 20 \;\mu$s to decode a QEC round, but it was evaluated on a Tensor Processing Unit (TPU).

\setcounter{lastequationbeforesection}{\number\value{equation}}
\section{Soft Information Processing} \label{ap_sec:prob_defects}
\setcounter{equation}{\number\value{lastequationbeforesection}}

The processing of the soft information [i.e., measurement edge weights in \Cref{eq:softedgeweight}, the defect probabilities in \cref{app_eq:d_prob}, and leakage flags], all use the probability density functions $P(z \mid j)$ which can be found from experimental calibration. These PDFs are obtained by fitting a readout model to the readout calibration data, consisting of a set of IQ values for each prepared state $\ket{j}$. The performance of the soft decoders is limited by the accuracy of the readout models used, thus in this section we describe the models employed in this paper and their underlying assumptions about the qubit readout response.

\subsection{Probability density function fits} \label{ap_sec:classifiers}

Each of the 13 transmons in the device has a characteristic measurement response in IQ space, requiring a unique PDF to be fitted for each. In this section, we detail a heuristic model used to classify qubit states (\cref{ap_sec:gaussian_mixture_model}) and qutrit states (\cref{ap_sec:3state-class}), formulated as a linear combination of Gaussian distributions. We utilise this Gaussian mixture model instead of a physics-derived measurement model such as the soft amplitude damping model derived in \cite{pattison2021improved} or the Bayesian approach taken in Ref.~\cite{cosco2023enhancing}. This is because we want to exclude the contribution of ancilla qubit errors that occur during measurement from the classification error probability. A Gaussian mixture fit allows us to classify states according to noisy experimental data, while excluding components representing $\ket{1} \rightarrow \ket{0}$ decay from the measurement error edge weights. It is also easily extended to incorporate measurements with leakage to the $\ket{2}$-state.

\subsubsection{Two-state Gaussian mixture model}\label{ap_sec:gaussian_mixture_model}

For discrimination between $\ket{0}$ and $\ket{1}$, we can use projected coordinates $\tilde{z}$ as defined in the main text, where the PDFs $P(\tilde{z} \mid j)$ for $j\in \{0, 1\}$ have the form

\begin{equation}\label{eq:1D_gaussian_mix_fit}
    P(\tilde{z} \mid j) = (1-r_j)f(\tilde{z}; \tilde{\mu}_0, \sigma) +r_j f(\tilde{z}; \tilde{\mu}_1, \sigma).
\end{equation}
\noindent Here, $f(\tilde{z}; \tilde{\mu}_i, \sigma)$ is a 1D Gaussian distribution with mean $\tilde{\mu}_j$ and standard deviation $\sigma$, and $r_j\in[0, 1]$ is an amplitude parameter that determines which normal distribution is dominant in the mixture. For $\{r_0=0, r_1=1\}$ the model represents a readout response with a single dominant component (i.e. no state preparation errors), while $\{r_0>0, r_1<1\}$ represents a measurement response where, due to state preparation errors, there are two distinct components to the measurement response.

The Gaussian mixture model allows us to discard state preparation errors from the $P(\tilde{z} \mid 0)$ by fitting the parameters $\tilde{\mu}_0, \tilde{\mu}_1$ and $\sigma$ for $r_0$ from \cref{eq:1D_gaussian_mix_fit} and then setting $r_0=0$. This assumption holds on the condition that no $\ket{0} \rightarrow \ket{1}$ processes are present over the course of the measurement time -- if significant amplitude damping occurs over the course of the measurement, the PDF found using the Gaussian method is inaccurate. When comparing experimental results for logical fidelity with and without setting $r_0=0$ for the ground state distribution $P(\tilde{z}\mid 0)$ we find no statistically significant difference. We assume the absence of a fidelity improvement is due to the rarity of $\ket{0}$ state preparation errors which are mitigated by heralded initialization. As mentioned in the main text, whilst we include both Gaussians in the PDF in order to classify the measurement, we use only the main peak in calculating the soft edge weights. This removes the component of qubit error that occurs during measurement from the edge associated with measurement classification error.

\begin{figure*}[t]
    \centering
    \includegraphics[width=380pt]{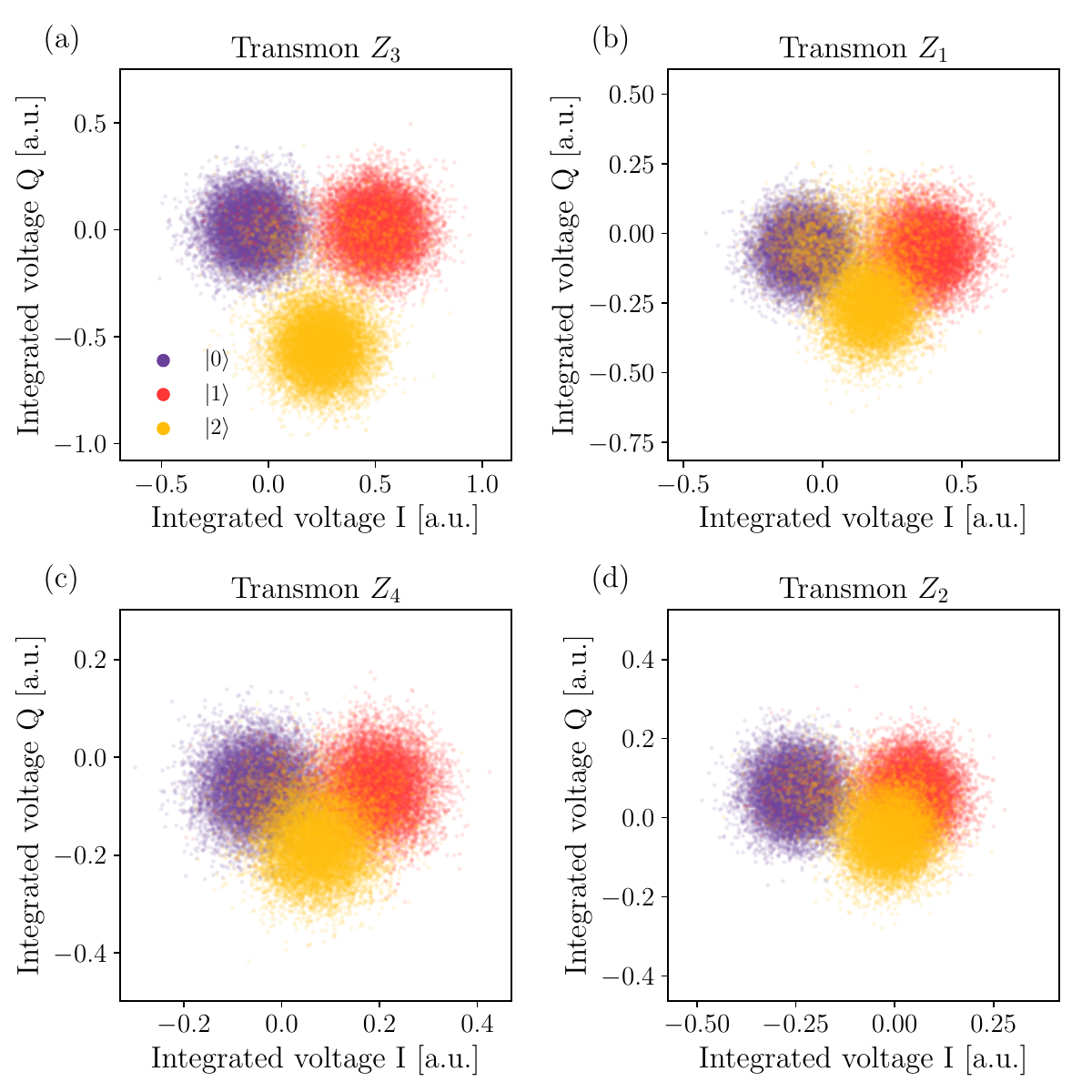}
    \caption{Calibration shots for experimental IQ voltages, shown for each ancilla prepared in $\ket{0}$, $\ket{1}$ and $\ket{2}$. The cluster for $\ket{2}$ is off-axis when compared to the clusters for $\ket{0}$ and $\ket{1}$, and its separation from the two other clusters varies from transmon to transmon. State $\ket{0}$ has the cleanest response, as it does not decay to any other state, while $\ket{1}$ and $\ket{2}$ decay partially to lower energy states over the course of measurement.}
    \label{ap_fig:2D_IQ_ancillas}
\end{figure*}

\subsubsection{Three-state classifier}
\label{ap_sec:3state-class}

The 1D projected model is unable to characterize leakage to $\ket{2}$, which has its own characteristic response in the two-dimensional IQ space, shown in \cref{ap_fig:2D_IQ_ancillas}. To model this three-state regime and discriminate leakage, we fit a mixture of 2D Gaussians to normalized histograms of the calibration data, giving PDFs $P(z \mid j)$ for $z \in \mathbb{R}^2$, $j \in \{0, 1, 2\}$ as follows:
\begin{equation}\label{eq:2D_gaussian_mix_fit}
    P(z \mid j) = A_j f(z; \vec{\mu}_0, \sigma) +B_j f(z; \vec{\mu}_1, \sigma) + C_j f(z; \vec{\mu}_2, \sigma),
\end{equation}
where $f(z; \vec{\mu}_j, \sigma)$ is the PDF of a 2D Gaussian distribution with mean $\vec{\mu}_j$ and covariance matrix $\sigma^2 I$, and parameters $A_j, B_j$, and $C_j$ are to be fitted for each state $\ket{j}$.

In \cref{ap_fig:2D_IQ_ancillas}, we observe that the $\ket{2}$ state has a measurement response that is off the $\vec{\mu}_0 - \vec{\mu}_1$ axis, forming a distinct constellation in the IQ space below the other two measurement responses $\ket{0}$ and $\ket{1}$. The ground state response is centred around $\vec{\mu}_0$, while the $\ket{1}$ response is distributed between a dominant peak around $\vec{\mu}_1$ and a small number of data points closer to $\vec{\mu}_0$, indicating $\ket{1} \rightarrow \ket{0}$ decay. The response of the $\ket{2}$ state can be seen to decay to both $\ket{0}$ and $\ket{1}$ states, most notably for transmon $Z_1$ where the effect can be clearly seen.

Given this simple model, the three-state classifier used to set the leakage flags as input for the NN decoder works as follows. Maximum likelihood classification means that given $z$, we should pick $j=0,1,2$ which maximizes $P(j|z)$ in \cref{app_eq:prob_k_z}. The denominator in \cref{app_eq:prob_k_z} can be dropped as it solely depends on $z$. For the numerator, we need to know $P(j)$ which we assume to be independent of $j$ (which is not completely warranted as $j=2$ is much less likely), and hence $P(j|z) \propto P(z\mid j)$ with $P(z\mid j)$ in \cref{eq:2D_gaussian_mix_fit}. 
These arguments identically apply to the two-state classifier discussed in the main text.

\subsection{Combining soft information with the pairwise correlation method in the final round}
\label{app:softinfo}
As discussed in the main text, we obtain the decoding graph edge weights from experimental data. These weights will include some averaged probability of a classification error that we wish to remove and replace with a soft-information-based weight on a per-shot basis. In the bulk of the experiment, this is straightforward -- the weight of the edge corresponding to a classification error can simply be replaced with that calculated using the soft information, following \Cref{eq:softedgeweight}, as classification and qubit errors have different defect signatures.

However, in the final round, both ancilla and data qubit errors result in the same defects as classification errors. Therefore, we expect the total (averaged across shots) edge probability to be given by
\begin{equation}
    p = q(1 - c) + c(1 - q), \label{eq:pqc}
\end{equation}
where $c$ is the averaged probability of any classification error and $q$ is the probability of qubit errors. The validity of this assumption depends on the degree to which we have made correct assumptions about the possible error channels, including that there is no correlated noise. In our case, $p$ is obtained by the pairwise correlation method, but it could be obtained by other means. We note that, in general, errors on multiple qubits may result in the same defects and thus contribute to the same edge probability.  In our Surface-13 experiment, this only occurs for certain final-round data qubit measurements -- the pairs D1 and D2, and D8 and D9.

We wish to retain the edge contributions due to the qubit errors, $q$, and replace only the averaged classification contribution, $c$, with the per-shot value. We thus have several steps to calculate the soft-information-based weight for final-round edges:
\begin{enumerate}[(i)]
    \item For each measurement $k$, calculate the mean classification error, $c^{k}$, by averaging the per-shot errors.
    \item For each edge, calculate the total edge classification error, $c$, from the individual measurement classification errors, $c^{k}$, through
    \begin{equation}
        c = \frac{1}{2} \left [ 1 - \prod_{^{k}}(1 - 2{c}^{k})\right ],
        \label{eq:cfromck}
\end{equation}
    where the product is over all measurements whose classification errors result in the edge defects.
    \item For each edge, remove the mean classification error from the edge probability by rearranging~\cref{eq:pqc} to find
    \begin{equation}
        q = \frac{p - c}{1 - 2c}.
        \label{eq:qfrompc}
    \end{equation}
    \item For each edge, include the per-shot classification error calculated from the soft readout information by combining it appropriately with $q$.
\end{enumerate}

We now explain the above steps in more detail.

\subsubsection{Step (i): calculating the mean classification error for each measurement}
For each experiment shot $s$, we have a soft measurement outcome for each measurement $k$, given by $z^{k}_{s}$, and associated inferred state after measurement, $\hat{z}'^{k}_{s}$. As discussed in the main text, in our case, this is found by taking $\hat{z}'^{k} = 1$ if $P'(z^{k} \mid 1) > P'(z^{k} \mid 0)$ and 0 otherwise. The PDFs $P'(z^{k} \mid j'^{k})$ are obtained by keeping only the dominant Gaussian in the measurement classification PDFs. We can calculate the estimated averaged probability of a classification error in measurement $k$, $c^{k}$, via
\begin{align}
    &c^{k} = \frac{1}{N}\sum_{s=1}^{N} {c}^{k}_{s} \\ &= \frac{1}{N}\sum_{s=1}^{N} \frac{P(1 - \hat{z}'^{k}_{s})P'(z^{k}_{s} \mid 1-\hat{z}'^{k}_{s})}{P( \hat{z}'^{k}_{s})P'(z^{k}_{s} \mid \hat{z}'^{k}_{s}) + P(1 - \hat{z}'^{k}_{s})P'(z^{k}_{s} \mid 1 - \hat{z}'^{k}_{s})},
\end{align}
where $P(j)$ is the overall probability, across all shots, that the qubit is in state $|j\rangle$; for example, if $\hat{z}'^{k}_{s} = 0$, then $P(\hat{z}'^{k}_{s}) = P(0)$. In this work, we take $P(0) = P(1) = 1/2$. Therefore, we have
\begin{equation}
    {c}^{k} = \frac{1}{N}\sum_{s=1}^{N} \frac{P'(z^{k}_{s} \mid 1-\hat{z}'^{k}_{s})}{P'(z^{k}_{s} \mid \hat{z}'^{k}_{s}) + P'(z^{k}_{s} \mid 1 - \hat{z}'^{k}_{s})}.
    \label{eq:ck}
\end{equation}

\subsubsection{Step (ii): calculating the mean classification error for each edge}
Using the estimated values for $c^{k}$ obtained with \cref{eq:ck}, we use \cref{eq:cfromck} to obtain the mean classification error, $c$, for each edge. In the case where an edge is only due to a single measurement classification error, we simply have $c = c^{k}$, for the relevant measurement $k$. The only other case of relevance for Surface-13 is where two classification errors contribute to an edge; in this case, $c = c^{k_{1}}(1 - c^{k_{2}}) + (1 - c^{k_{1}})c^{k_{2}}$ for the relevant measurements $k_{1}$ and $k_{2}$. 

\subsubsection{Step (iii): removing the mean classification error from each edge probability}
We now calculate $q$ for each edge using \cref{eq:qfrompc}.

\subsubsection{Step (iv): including the per-shot classification error}
We now wish to combine the per-shot soft information with $q$ in order to obtain the full edge weight. In doing so, we mostly follow Ref.~\cite{pattison2021improved}, with the difference that we merge edges that result in the same defects into a single edge. In everything that follows, we are considering a single experiment shot and thus, to reduce notational clutter, drop the index $s$ from above.

We begin by considering the full decoding problem we wish to solve. We have a set of possible errors $\mathbb{E}$ and consider a single error, $e_{i}$, to consist of all events that contribute to the same edge. This includes both classification errors and other errors. We further have a set of (labelled) soft measurement outcomes, $\mathbb{Z}$, which is the union of all sets $\mathbb{Z}_{i}$, where $\mathbb{Z}_{i}$ is the set of measurements whose incorrect classification leads to the same defect combination as $e_{i}$. We wish to find the combination of errors $\mathbb{D}$ that explains the observed defects and maximizes
\begin{equation}
P(\mathbb{D} \mid \mathbb{Z}) = \frac{P(\mathbb{D} \cap \mathbb{Z})}{P(\mathbb{Z})} \propto P(\mathbb{D} \cap \mathbb{Z}),
\end{equation}
where $\mathbb{D} \cap \mathbb{Z}$ is the event that the combination of errors $\mathbb{D}$ occur and the soft measurements outcomes $\mathbb{Z}$ are obtained. We can ignore the denominator $P(\mathbb{Z})$ as it is a constant rescaling of all probabilities $P(\mathbb{D}\mid \mathbb{Z})$ and thus does not need to be considered in order to find the most likely error.

Assuming independence of events, we split $P(\mathbb{D} \cap \mathbb{Z})$ into individual terms for each edge so that
\begin{equation}
    P(\mathbb{D} \cap \mathbb{Z}) = \prod_{e_{i} \in \mathbb{D}} P(e_{i} \cap \mathbb{Z}_{i}) \prod_{e_{i} \notin \mathbb{D}} P(\bar{e_{i}} \cap \mathbb{Z}_{i}).
\end{equation}
Rearranging, we find
\begin{align}
    P(\mathbb{D} \cap \mathbb{Z}) &= \prod_{e_{i} \in \mathbb{E}} P(\bar{e_{i}} \cap \mathbb{Z}_{i}) \prod_{e_{i} \in \mathbb{D}} \frac{P(e_{i} \cap \mathbb{Z}_{i})}{P(\bar{e_{i}} \cap \mathbb{Z}_{i})} \\ &\propto \prod_{e_{i} \in \mathbb{D}} \frac{P(e_{i} \cap \mathbb{Z}_{i})}{P(\bar{e_{i}} \cap \mathbb{Z}_{i})},
\end{align}
where, again, we can drop the term that is common to all error combinations. Maximizing $P(\mathbb{D} \cap \mathbb{Z})$ is equivalent to minimizing 
\begin{equation}
    - \log [P(\mathbb{D} \cap \mathbb{Z})] = - \sum_{e_{i} \in \mathbb{D}} \log \left[ \frac{P(e_{i} \cap \mathbb{Z}_{i})}{P(\bar{e}_{i} \cap \mathbb{Z}_{i})} \right ] \equiv \sum_{e_{i} \in \mathbb{D}} w_{i},
\end{equation}
where we have defined
\begin{equation}
    w_{i} = - \log \left[ \frac{P(e_{i} \cap \mathbb{Z}_{i})}{P(\bar{e}_{i} \cap \mathbb{Z}_{i})} \right ].
    \label{eq:weight}
\end{equation}
Let us now consider a particular error, $e_{i}$, and its $n_{i}$ associated soft measurements $z_{i}^{k}$ for $k=1, \dots, n_{i}$. We recall that, in the Surface-13 case, $n_{i}$, which is the number of classification errors that contribute to edge $i$, is a maximum of two, and we use this below. In order to calculate $w_{i}$, we split $e_{i}$ into two: $e_{i}^{c}$, which consists of classification errors only, and $e_{i}^{q}$ which consists of all other errors. There are now two ways in which $e_{i}$ can occur: (i) $e_{i}^{q}$ occurs and $e_{i}^{c}$ does not occur (i.e. there are an even number of classification errors); (ii) $e_{i}^{q}$ does not occur and $e_{i}^{c}$ does occur (i.e. there are an odd number of classification errors). Therefore,
\begin{align*}
    P(e_{i} \cap \mathbb{Z}_{i}) &= P(e_{i}^{q})P(\bar{e}_{i}^{c} \cap \mathbb{Z}_{i}) + P(\bar{e}_{i}^{q})P(e_{i}^{c} \cap \mathbb{Z}_{i}) \\ &= qP(\bar{e}_{i}^{c} \cap \mathbb{Z}_{i}) + (1 - q)P(e_{i}^{c} \cap \mathbb{Z}_{i}), \\
    P(\bar{e}_{i} \cap \mathbb{Z}_{i}) &= P(e_{i}^{q})P(e_{i}^{c} \cap \mathbb{Z}_{i}) + P(\bar{e}_{i}^{q})P(\bar{e}_{i}^{c} \cap \mathbb{Z}_{i}) \\ &= qP(e_{i}^{c} \cap \mathbb{Z}_{i}) + (1 - q)P(\bar{e}_{i}^{c} \cap \mathbb{Z}_{i}),
\end{align*}
where we have defined $P(e_{i}^{q}) = q$. 

\begin{widetext}
The probability of obtaining the observed soft measurement outcomes and having an \textit{odd} number of classification errors is
\begin{equation}
P(e_{i}^{c} \cap \mathbb{Z}_{i}) =
\begin{cases}
    P'(z_{i}^{1} \mid 1-\hat{z}_{i}'^{1}), & n_{i}=1, \\ \nonumber
    P'(z_{i}^{1} \mid 1-\hat{z}_{i}'^{1})P'(z^{2}_{i} \mid \hat{z}_{i}'^{2}) + P'(z^{1}_{i} \mid \hat{z}_{i}'^{1})P'(z^{2}_{i} \mid 1-\hat{z}_{i}'^{2}), & n_{i}=2,
\end{cases}
\end{equation}
and the probability of obtaining the observed soft measurement outcomes and having an \textit{even} number of classification errors is

\begin{equation}
P(\bar{e}_{i}^{c} \cap \mathbb{Z}_{i}) =
\begin{cases}
    P'(z_{i}^{1} \mid \hat{z}_{i}'^{1}), & n_{i}=1, \\
    P'(z_{i}^{1} \mid \hat{z}_{i}'^{1})P'(z^{2}_{i} \mid \hat{z}_{i}'^{2}) + P'(z^{1}_{i} \mid 1-\hat{z}_{i}'^{1})P'(z^{2}_{i} \mid 1-\hat{z}_{i}'^{2}), & n_{i}=2.
\end{cases}
\end{equation}
These expressions can easily be extended to larger values of $n_{i}$, but we omit the general expressions here for brevity. 
\end{widetext}
From these, we calculate the edge weight using \Cref{eq:weight} and find
\begin{equation}
    w_{i} = 
    \begin{cases}
        - \log \left(\frac{L_{i}^{1} + L_{i}^{q}}{1 + L_{i}^{1}L_{i}^{q}} \right ), & n_{i}=1, \\
        - \log \left(\frac{L_{i}^{1} + L_{i}^{2} + L_{i}^{q} + L_{i}^{1}L_{i}^{2}L_{i}^{q}}{1 + L_{i}^{1}L_{i}^{2} + L_{i}^{1}L_{i}^{q} + L_{i}^{2}L_{i}^{q}} \right ), & n_{i}=2,
    \end{cases}
\end{equation}
where
\begin{align}
    L_{i}^{q} &= \frac{q}{1 - q}, \\
    L_{i}^{k} &= \frac{P'(z^{k}_{i} \mid 1-\hat{z}_{i}'^{k})}{P'(z^{k}_{i} \mid \hat{z}_{i}'^{k})}.
\end{align}

\begin{figure}[t]
    \centering
    \includegraphics[width=0.5\textwidth]{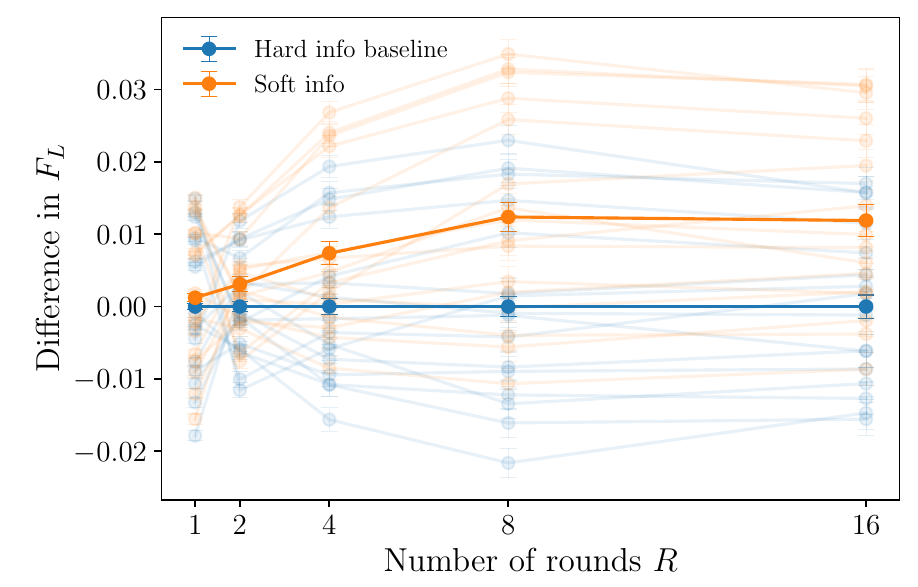}
    \caption{Absolute difference in logical fidelity $F_L$ as a function of rounds, shown for each individual logical state preparation $\ket{0_{\mathrm{L}}}$ (transparent) and on average across 16 states (opaque) for soft versus hard MWPM.}
    \label{fig:ler_difference_soft_mwpm}
\end{figure}

\setcounter{lastequationbeforesection}{\number\value{equation}}
\section{Calculation of Logical Error Rate}\label{ap_sec:logical_performance}
\setcounter{equation}{\number\value{lastequationbeforesection}}

To extract the logical error rate $\epsilon_{\mathrm{L}}$ from experimental data, we calculate the logical fidelity $F_{\mathrm{L}}(R)$ for each round $R$ of the experiment and fit the data to a decay curve of the form given in \Cref{eq:logical_fidelity}. The error in the logical fidelity is given by $\sigma^2_{F_{\mathrm{L}}}=F_{\mathrm{L}}(1 - F_{\mathrm{L}}) / N$, with $N$ the number of samples for the given $F_{\mathrm{L}}$ \cite{bausch2023learning}, which we propagate through the fitting process to get estimates of uncertainty in $\epsilon_{\mathrm{L}}$ and the offset $R_0$.

\setcounter{lastequationbeforesection}{\number\value{equation}}
\section{Additional Logical Error Rate Figures}
\setcounter{equation}{\number\value{lastequationbeforesection}}

\begin{figure}
    \centering
    \includegraphics[width=0.4\textwidth]{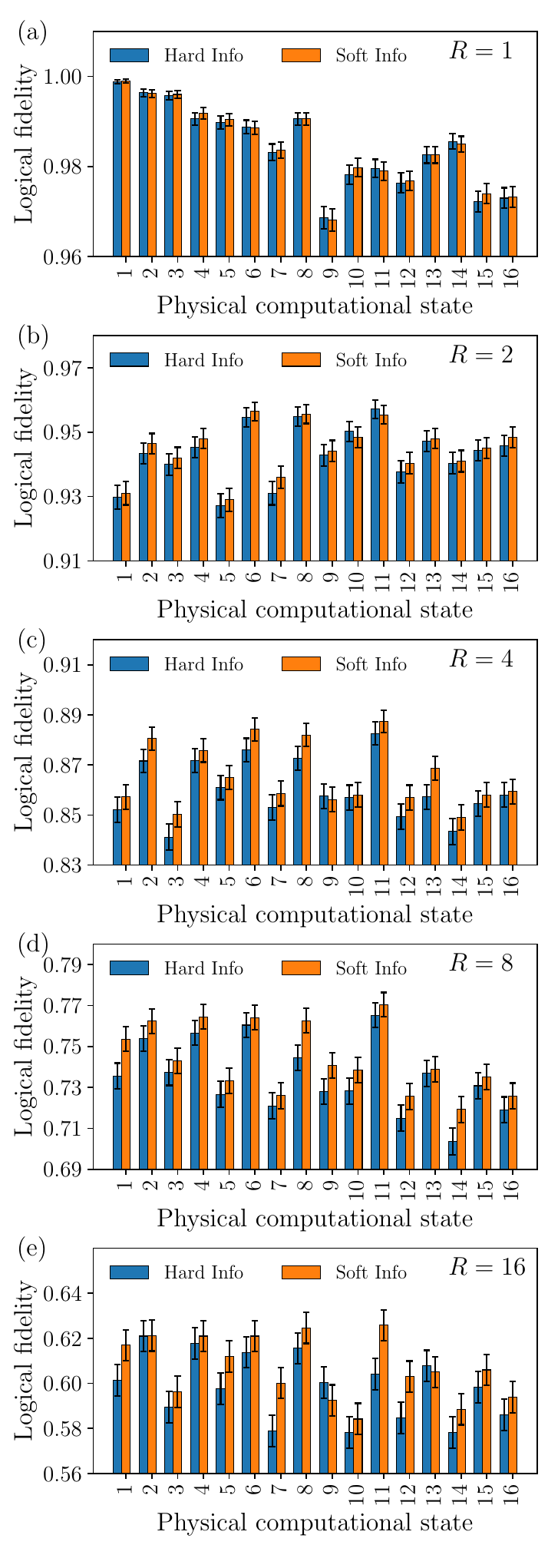}
    \caption{Logical fidelity using soft versus hard MWPM decoder for each round of the experiment.}
    \label{fig:LER_bar_all_rounds_mwpm}
\end{figure}

We show additional plots of the logical fidelity of the soft and hard MWPM decoders for each round of the experiment in \cref{fig:LER_bar_all_rounds_mwpm}. To illustrate the improvement that soft information gives to logical fidelity, \cref{fig:ler_difference_soft_mwpm} shows the absolute difference in logical fidelity $F_{\mathrm{L}}(R)$ between the soft and hard MWPM decoders for each round of the experiment. The average performance is shown in solid lines, and the fidelity for each individual state preparation $\ket{0_{\mathrm{L}}}$ is shown in the transparent lines.

\newpage

% \bibliography{references}

%apsrev4-2.bst 2019-01-14 (MD) hand-edited version of apsrev4-1.bst
%Control: key (0)
%Control: author (72) initials jnrlst
%Control: editor formatted (1) identically to author
%Control: production of article title (-1) disabled
%Control: page (0) single
%Control: year (1) truncated
%Control: production of eprint (0) enabled
%

\end{document}